\documentclass[fleqn,usenatbib]{mnras}
\usepackage{graphicx}
\usepackage{epstopdf}

\newcommand\un[1]{{\,\rm #1}}
\newcommand\E[1]{\times10^{#1}}
\newcommand\rs[1]{_\mathrm{#1}}
\newcommand\g{$\gamma$}

\usepackage[T1]{fontenc}
\usepackage{ae,aecompl}

\title[Post-adiabatic SNRs in ambient magnetic field]{Post-adiabatic supernova remnants in the interstellar magnetic field. 
Parallel and perpendicular shocks}
\author[O. Petruk, T. Kuzyo, V. Beshley]{O. Petruk, T. Kuzyo, V. Beshley\\
Institute for Applied Problems in Mechanics and Mathematics, Naukova St 3-b, 79060 Lviv, Ukraine
}

\date{Last updated ...; in original form ...}
\pubyear{2015}

\begin{document}

\label{firstpage}
\pagerange{\pageref{firstpage}--\pageref{lastpage}} 
\maketitle

\begin{abstract}
Gamma-rays from hadronic collisions are expected from supernova remnants (SNRs) located near molecular clouds. The temperature on the shock interacting with the dense environment quickly reaches $10^5$ K. The radiative losses of plasma become essential in the evolution of SNRs. They decrease the thermal pressure and essentially increase the density behind the shock. The presence of ambient magnetic field may considerably alter the behavior of the post-adiabatic SNRs comparing to hydrodynamic scenario. In the present paper, the magneto-hydrodynamic simulations of radiative shocks in magnetic field are performed. High plasma compression due to the radiative losses results also in the prominent increase of the strength of the tangential component of magnetic field behind the shock and the decrease of the parallel one. If the strength of the tangential field before the shock is higher than about $3\un{\mu G}$ it prevents formation of the very dense thin shell. The higher the strength of the tangential magnetic field the larger the thickness and the lower the maximum density in the radiative shell. Parallel magnetic field does not affect the distribution of the hydrodynamic parameters behind the shock. There are almost independent channels of energy transformations: radiative losses are due to the thermal energy, the magnetic energy increases by reducing the kinetic energy. The large density and high strength of the perpendicular magnetic field in the radiative shells of SNRs should result in considerable increase of the hadronic gamma-ray flux comparing to the leptonic one.
\end{abstract}


\section{Introduction}

Ground- and space-based observations of \g-rays from supernova remnants (SNRs) prove that galactic cosmic rays are produced in these objects. It is shown that \g-rays are likely leptonic in RX J1713.7-3946 \citep{Abdo-et-al-2011} and hadronic in IC 443 and W44 \citep{Ackermann-et-al-2013}. In most other SNRs, \g-rays remain of unknown origin because theoretical models may explain observed \g-ray spectra in both frameworks: either with relativistic  electrons or with protons \citep[e.g.][]{SN1006-HESS}. There are some hints, both theoretical \citep{Petruk-et-al-2009} and observational \citep{Acero-et-al-2015}, that SNRs having the shell-like morphology in \g-rays are likely emitting through the leptonic mechanism.

There are many evidences that electrons are accelerated in SNRs from observations in radio and nonthermal X-rays. Instead, direct observational proofs about acceleration of protons in SNR shocks are difficult to obtain because of small radiative ability of these particles. Photons emitted in hadronic collisions may be observed in \g-rays only. However, photons from inverse-Compton interactions with electrons have the same energies. Hadronic \g-rays should dominate in systems with rather high magnetic field (in order to reduce efficiency of leptonic \g-rays) and with high density of cold protons which are targets for relativistic protons accelerated by the SNR shock. 

Such conditions naturally appear when the shock moves in the medium with the high density, e.g. it enters into a molecular cloud. Therefore, SNRs located near molecular clouds are promi\-sing sites to observe hadronic \g-rays \citep{Slane-et-al-2014}. Thus, there is a reason for a systematic search for such SNRs \citep{Jiang-et-al-2010,Jiang-et-al-2014}. 

The shock in SNR interacting with the molecular cloud decelerates because of increased density. As a result, the temperature of the post-shock plasma decreases and radiative losses become important. The shock gradually transforms from the adiabatic to the radiative one \citep[and references therein]{Blondin-et-al-98}. The physics of the shock and conditions for particle acceleration change considerably. 

Picture of SNR evolution consists typically of the three stages -- free expansion, adiabatic and radiative -- each with its own features. Dissipation of very old SNRs could be considered as the fourth stage \citep[e.g.][]{Slavin-Cox-1992}. It is often assumed that these phases change one another very rapidly comparing to the duration of the phases themselves. However, the transition from the fully adiabatic to the fully radiative shock lasts almost the same time as duration of the adiabatic stage \citep{petruk2005}. It has its own, different from other stages, features \citep{Blondin-et-al-98}. First, the presence of the dynamically important radiative losses does not allow the \citet{Sedov-59} model to be used. Second, the thin cold radiative shell is not yet formed and, therefore, the solution for the radiative shock \citep{McKee-Ostr-1977,Pasko-Silich-86,Band-Pet-2004} is not relevant. 
Therefore, there is the need to separately consider the post-adiabatic stage in a general scenario of the SNR evolution, especially in SNRs interacting with molecular clouds.

Evolution of SNR during the post-adiabatic stage is accompanied by increasing role of the radiative losses and therefore essential change of the flow structure. In particular, the regions with high density and magnetic field (MF) should appear right after the shock. It is expected that such flow reconstruction has to result in the prominent increase of the hadronic \g-rays (due to the high density of target protons) and the decrease of the lepronic \g-rays (due to synchrotron losses of electrons in high MF). Thus, the post-adiabatic SNRs could be sources of the hadronic emission. 

Hydrodynamic properties of the post-adiabatic flows are demonstrated in the numerical simulations of \citet{Cioffi-et-al-88,Blondin-et-al-98}. \citet{Hn-Pet-Tel-2007} have developed an approximate semi-analytical method to describe the radiative shock without MF. It bases on the main properties of the flow during transition from the adiabatic to the fully radiative regime and neglects minor effects. This hydrodynamic method is applied to a study of the hadronic \g-ray emission from SNRs by \citet{Tel-Hn-2007}. 

Magnetic field becomes an important factor of the SNR evolution just in the post-adiabatic SNRs. Radiative losses lead to increase of the shock compression and thus to increase of the MF strength. The energy density of MF becomes  comparable to the thermal energy density. MF affects dynamics of the shock and the flow downstream \citep{Falle-1975,Slavin-Cox-1992,Innes-1992}. In particular, MF limits the density of the radiative shell. In the simulations without MF, the ratio of the post-shock to the pre-shock density may reach few hundred \citep{Blondin-et-al-98}. 

In the present paper, the evolution of the post-adiabatic SNRs in the interstellar medium with magnetic field is studied. We solve numerically the system of differential equations of the ideal MHD in order to see how does MF affect the evolution of the post-adiabatic SNRs. 

Note, that other factors could influence the formation and compression of the radiative shell, namely, thermal conduction \citep{Slavin-Cox-1992,Orlando2008} and cosmic ray pressure in case of efficient acceleration \citep{Lee-et-al-2015}. They are outside of the scope of the present paper.

\section{Preliminary simulations}

\subsection{The code and problem setup}

The numerical code PLUTO \citep{Mignone2007,Mignone2012} is adopted for our simulations. It is designed to describe the supersonic flows in the presence of strong shocks. PLUTO integrates the system of the time-dependent conservation laws of the ideal MHD in the form
\begin{equation}
 \frac{\partial}{\partial t} \left(\!\!
  \begin{array}{c}
  \rho \\ \mathbf{m} \\ E \\ \mathbf{B}
  \end{array}
 \!\!\right) + \nabla \cdot \left(\!\!\!
  \begin{array}{c}
  \rho \mathbf{v} \\
  \mathbf{m} \mathbf{v} + \mathbf{I}p\rs{tot} \\
  (E+p\rs{tot})\mathbf{v} - \mathbf{B}(\mathbf{v}\cdot\mathbf{B}) \\
  \mathbf{v}\mathbf{B} - \mathbf{B}\mathbf{v} \\
 \end{array} 
 \!\!\!\right)^\mathrm{T} \!\! = \left(\!\!
  \begin{array}{c}
  0 \\ 0 \\ L \\ 0
 \end{array}
 \!\!\right)
\label{postadiab:eq2}
\end{equation}
where $\rho$ is the density, $\mathbf{m}=\rho \mathbf{v}$ the momentum density, $\mathbf{v}$ the flow velocity, $p\rs{tot}$ the total (thermal $p$ and magnetic $p\rs{B}$) pressure, $\mathbf{I}$ the unit vector, $\mathbf{B}$ the MF strength, $L$ represents the radiative losses, $E$ the total energy density. 
The ideal gas equation of state is assumed with $\gamma=5/3$. The total energy density is a sum of the thermal, kinetic and magnetic components:
\begin{equation}
 E=\frac{p}{\gamma-1}+\frac{m^2}{2\rho}+\frac{B^2}{2}.
\label{postadiab:baseeq}
\end{equation}
Additionally, the divergence-free condition holds: $\nabla \cdot \mathbf{B}=0$.

The system (\ref{postadiab:eq2}) is integrated with the use of the finite volumes approach.
The solving strategy in PLUTO consists of the three main steps:
polynomial interpolation of the cell-averaged values to the interfaces of cells,
solving the Riemann problem at the cell edges and the time evolution.
For each of the steps several possible algorithms are available in the code.
Therefore, one can set up a number of computational schemas. We have tried
a series of different combinations of available alternative algorithms
in order to choose the one which seems to be the most appropriate for our goals.
Namely, we used the following set in our simulations: {\sf linear} interpolation with {\sf min mod}
limiter, {\sf HLL} Riemann solver and {\sf Characteristic Tracing} algorithm for the time evolution.
Additionally, we used {\sf oned} shock flattening algorithm for the numerical
dissipation near the strong shocks and {\sf eight wave} formulation for controlling
$\nabla \cdot \mathbf{B}=0$ condition.
Such numerical setup provides the second order accuracy both in spatial and
temporal integrations.

Our one-dimensional spherically-symmetrical computations were carried on the
static uniform grid. The physical grid size is 32 pc with 60 000 computational 
zones\footnote{Other resolutions were tried, namely,  10\,000, 20\,000 and 30\,000 zones, but the strong dense peaks  from simulations of \citet{Blondin-et-al-98} were correctly reproduced in the hydrodynamical simulations with 60\,000 zones.}.
We used the reflective boundary conditions (variables are symmetric across the boundary
and the normal components of vectors are antisymmetric) at the beginning of the grid as well as 
the outflow (zero gradient across the boundary) at the end of the grid.
The time step was limited by the CFL condition with Courant number 0.75.
The simulations were carried until $t = 100\,000\un{yrs}$. 

There are known solutions of the system (\ref{postadiab:eq2}). Namely, \citet{Sedov-59} found the analytical solutions for $L=0$ and $B=0$ \citep[for approximations of the Sedov solutions see][]{approxSedov}. Analytical description of MF in Sedov SNRs ($L=0$, $B\neq0$) are given by \citet{Korob1960,Korob1964}. 
The numerical solutions for $L\neq 0$ and $B=0$ are presented by \citet{Cioffi-et-al-88,Blondin-et-al-98}. 

The goal of the present paper is to study the case of the post-adiabatic shocks in the ambient MF, i.e. $L\neq0$ and $B\neq 0$. The spherically symmetric problem is considered.

\begin{figure}
 \centering
 \includegraphics[width=8.4truecm]{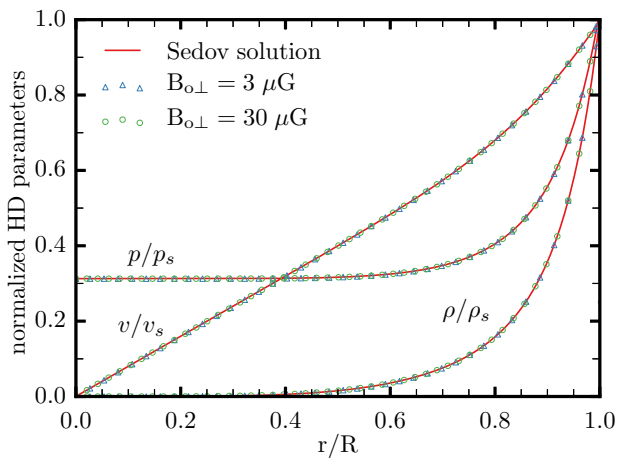}
 \caption{Distribution of HD parameters behind the perpendicular shock. Sedov solutions (without MF; solid lines) are compared with MHD numerical simulations (open squares and triangles). Data are presented for two values of ISMF strength, $n\rs{o}=0.3\un{cm^{-3}}$, $t=1000\un{yrs}$. Plots for other values of $t$ and the parallel shock are the same.}
 \label{vb-fig2a}
\end{figure}

\subsection{Magnetic field in Sedov SNRs}

Initially, the hydrodynamic (HD) module in the PLUTO code was tested on \citet{Sedov-59} analytical solutions for uniform ISM. The self-similarity of the spatial profiles of the hydrodynamic parameters (density, pressure and velocity) was reproduced in the numerical simulations. 

On the next step, MHD simulations were carried out of the adiabatic SNR in uniform ISM and uniform ISMF in order to reveal the influence of the magnetic field on the evolution of HD parameters behind the shock as well as to determine the postshock evolution of magnetic field. 

The simulations demonstrate (Fig.~\ref{vb-fig2a}) that MF may be considered as `test-like', i.e. it is not capable to change the distribution of the hydrodynamic parameters inside the adiabatic SNR. The reason is that the energy density of MF (with the strengths which are expected in SNRs) is much smaller than the thermal energy density of the gas (high plasma $\beta$). Thus, in order to model the MF inside the adiabatic SNR one could separate MF and hydrodynamics: one should first know the HD structure and then, on this hydrodynamic background, one may calculate the MF distribution. 

Actually, such an approach is already adopted in a number of publications in order to develop an analytical description of MF in Sedov SNR. Note, that the strong shock from the point explosion is spherical even in the presence of ambient MF, because the deviation from the sphericity $\propto M\rs{A}^{-2}\ll 1$ \citep[][p.201]{fleishman-topt-2013}. The whole problem is axisymmetric if the ambient MF is uniform. 

The changes of the uniform MF due to the strong point explosion in the ideal plasma with infinite conductivity and `weak' MF are considered by \citet{Korob1960}. The equations
\begin{equation} 
 \frac{\partial \mathbf{B}}{\partial t}=\mathrm{rot} [\mathbf{v}\times \mathbf{B}] , \quad 
 \mathrm{div} \mathbf{B}=0
\end{equation}
transform to the equations for the radial $B\rs{\|}$ and tangential $B\rs{\perp}$ components of MF in the spherical coordinates
\begin{equation} 
 \frac{\partial {B\rs{\|}}}{\partial t}+\frac{v}{r^2}\frac{\partial }{\partial r} \left(r^2B\rs{\|}\right)=0, \quad 
 \frac{\partial {B\rs{\perp}}}{\partial t}+\frac{1}{r}\frac{\partial }{\partial r} \left(rvB\rs{\perp}\right)=0
 \label{rad:eqMFsedov_sphericla}
\end{equation}
where $r$ is the radius, $\theta$ the angle measured from the direction of $\mathbf{B}\rs{o}$. At the shock front,
\begin{equation} 
 B\rs{\|,s}=B\rs{\|,o}, \quad 
 B\rs{\|,o}=B\rs{o}\cos\theta,
\end{equation}
\begin{equation} 
 B\rs{\perp,s}=\sigma B\rs{\perp,o}, \quad
 B\rs{\perp,o}=B\rs{o}\sin\theta,
\end{equation}
where $\sigma=\rho\rs{s}/\rho\rs{o}$ is the shock compression ratio, the index `o' refers to the pre-shock position, the index `s' to the immediately post-shock one. The self-similar solution of Eq.~(\ref{rad:eqMFsedov_sphericla}) may be found, -- after substitution (\ref{rad:eqMFsedov_sphericla}) with the \citet{Sedov-59} solution for the flow velocity $v(r,t)=v\rs{s}(t)\bar v(\bar r)$, -- by separation of variables (overline denotes the value normalized by the value of the same parameter at the shock, e.g. $\bar r=r/R$, $\bar B\rs{\|}=B\rs{\|}/B\rs{\|,s}$, $\bar B\rs{\perp}=B\rs{\perp}/B\rs{\perp,s}$). The  expressions for the self-similar solution for MF are quite large \citep{Korob1960}. 

An approximation of this solution is found assumming that $\bar v(\bar r)\approx \bar r$ in Sedov solutions. If so, the downstream profiles of both MF components would have the same shape \citep{toptygin2004}:
\begin{equation} 
 \bar B\rs{\|}=\bar B\rs{\perp}\approx\bar r^{\ 4/(\gamma-1)}.
 \label{rad:appsolMF}
\end{equation}

\begin{figure}
 \centering
 \includegraphics[width=8.7truecm]{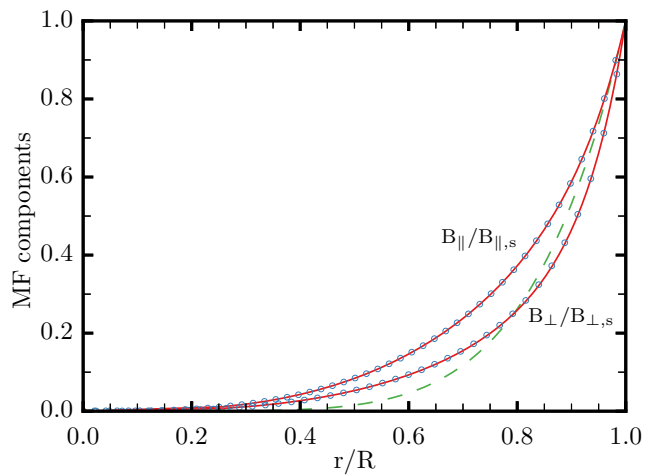}
 \caption{Radial ($B_\|$) and tangential ($B_\perp$) components of MF downstream of the Sedov shock. {Solid line} shows the analytic expressions (\ref{rad:MFlagr}), {dots} represent our numerical simulations. Approximate solution (\ref{rad:appsolMF}) is shown by the dashed line.}
 \label{vb-fig3}
\end{figure}
\begin{figure*}
 \centering
 \includegraphics[width=15truecm]{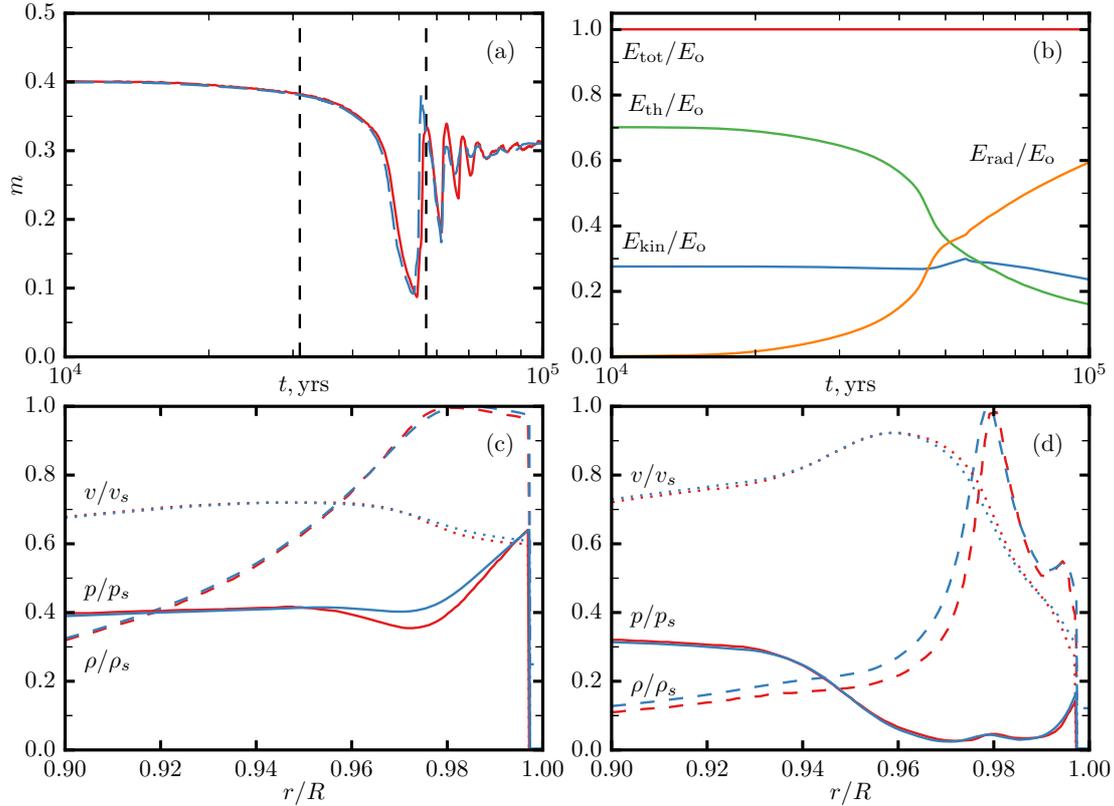}
 \caption{Expansion parameter $m=Vt/R$ ({\bf a}), the energy budget and its components ({\bf b}) and the post-shock distributions of HD parameters close to the shock front ({\bf c-d}) for SNR in uniform ISM without MF, with the radiative losses accounted for. Age of SNR is $t=44000\un{yrs}$ {\bf (c)} and $t=50000\un{yrs}$ {\bf (d)}. The vertical lines on plot ({\bf a}) represents the `transition' time $t\rs{tr}$ and the `shell formation' time $t\rs{sf}$.
Red lines on plots {(a), (c) and (d)} represent our calculations, the blue lines are for results of \citet{Blondin-et-al-98}. $v\rs{s}$, $p\rs{s}$ and $\rho\rs{s}$ are appropriate Sedov post-shock values.}
 \label{tk-fig5bl}
\end{figure*}

The above solutions are derived in terms of the Euler coordinates $r$. In many situations an analysis in terms of the Lagrangian coordinates $a$ is preferable. There are solutions of that problem as well \citep{Korob1964,Chevalier1974,Reyn-98} and they are much simpler:
\begin{equation} 
 \bar B\rs{\|}=\left(\frac{\bar a}{\bar r}\right)^2, \quad
 \bar B\rs{\perp}=\bar\rho\frac{\bar r}{\bar a}.
 \label{rad:MFlagr}
\end{equation}
These equations reflect the flux-conservation and freezing conditions. In order to have the dependence $B(a)$, one should use the known \citet{Sedov-59} relations for $r(a)$ and $\rho(a)$ or their much simpler approximations \citep[][Eqs.~86, 87]{approxSedov}. On the other hand, in order to have $B(r)$, one may use (\ref{rad:MFlagr}) with \citet{Sedov-59} relations for $a(r)$ and $\rho(r)$ or their approximations \citep[][Eqs.~A2, A5]{Cox-Franco-1981}.

The MF strength at any place in Sedov SNR is 
\begin{equation}
 B=B\rs{o}\left(\bar B\rs{\|}^2\cos^2\theta+\sigma^2 \bar B\rs{\perp}^2\sin^2\theta\right)^{1/2}. 
\end{equation}
The formula for the MF compression ratio at the shock for an arbitrary $\theta$ is therefore
\begin{equation}
 \sigma\rs{B}\equiv \frac{B\rs{s}}{B\rs{o}}=\left(\cos^2\theta+\sigma^2 \sin^2\theta\right)^{1/2}
\end{equation}
or \citep{Reyn-98}
\begin{equation}
 \sigma\rs{B}=
 \left(\frac{1+\sigma^2\tan^2 \theta}{1+\tan^2 \theta}\right)^{1/2}.
\end{equation}

Fig.~\ref{vb-fig3} shows that the analytical solutions (\ref{rad:MFlagr}) derived under assumption of the `weak' field are in good agreement with our numerical simulations which are free of such simplification. 

\subsection{Hydrodynamics of the post-adiabatic SNRs}

At some time of SNR evolution, radiative losses become high enough and the adiabatic condition no longer holds. This marks the end of the adiabatic stage. 

Radiative energy losses of the plasma are described with the term
\begin{equation}
 L=-n\rs{e}n\rs{H}\Lambda(T)
\end{equation}
where $T$ is the plasma temperature, $\Lambda(T)$ is the equilibrium \citep[e.g.][]{Raymond-Cox-Smith-1976} or nonequilibrium \citep[e.g.][]{Sutherland-Dopita-1993} cooling coefficient. 

We carried out the 1-D HD simulations ($B=0$) of the spherical SNR in the uniform ISM. In all simulations in the present paper, we use the same initial parameters as in \citet{Blondin-et-al-98}. Namely, the explosion energy $E\rs{o}=10^{51}\un{erg}$, the hydrogen number density in ISM $n\rs{H}=0.84\un{cm^{-3}}$, the temperature in ISM $10^4\un{K}$. The nonequilibrium cooling function $L$ is taken from \citet{Sutherland-Dopita-1993}. 

The results of \citet{Blondin-et-al-98} were reproduced if the mean mass per particle, per hydrogen atom, per free electron in the terms of the proton mass are $\mu=13/21$, $\mu\rs{H}=13/9$, $\mu\rs{e}=13/11$ respectively. This is demonstrated on Fig.~\ref{tk-fig5bl}. 

\begin{figure*}
\centering 
\includegraphics{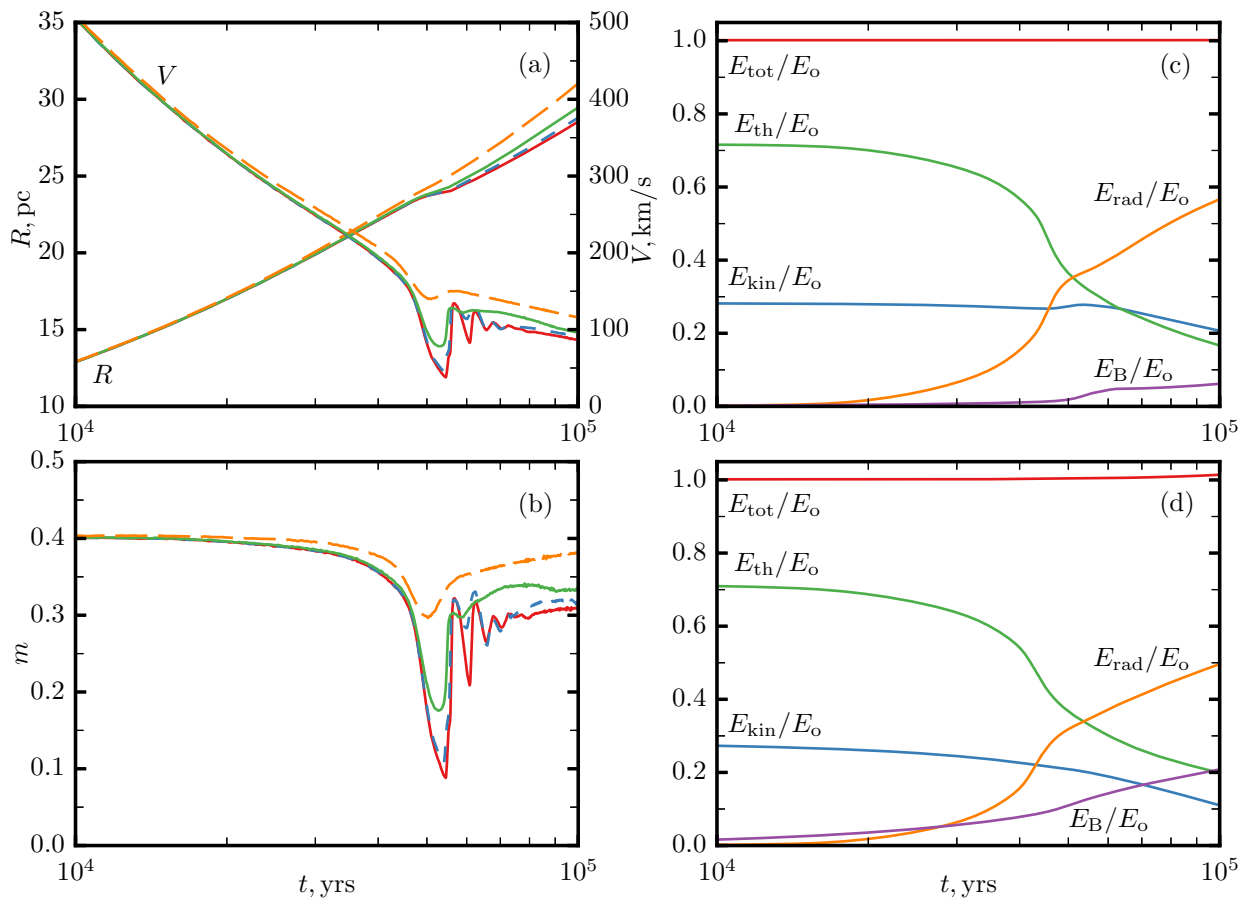}
\caption{Effect of magnetic field on the dynamics of shock with radiative losses. 
Time dependence of the shock radius $R$ and velocity $V$ ({\bf a}) as well as the expansion
parameter $m$ ({\bf b}) are shown for $B_{o} = 0\un{\mu G}$ (red line), $B_{o\perp} = 3\un{\mu G}$ (blue line), 
$B_{o\perp} = 10\un{\mu G}$ (green line) and $B_{o\perp} = 30\un{\mu G}$ (orange line). 
Plots for parallel shock coincide with $B_{o} = 0\un{\mu G}$ for any value of $B_{o\parallel}$.
Temporal evolution of the energy components are presented for $B_{o\perp} = 10\un{\mu G}$ ({\bf c}) 
and $B_{o\perp} = 30\un{\mu G}$ ({\bf d}).}
\label{fig_blast_dynamics}
\end{figure*}
	
Plot for the expansion parameter $m=-d\ln R/d\ln t=Vt/R$ (i.e. $R\propto t^{m}$) on Fig.~\ref{tk-fig5bl}a 
is useful in order to see the evolutionary stage of the shock. The deviation of $m$ from $0.4$ marks the end of the  adiabatic regime and beginning of the post-adiabatic phase. It happens around the transition time \citep{Cox-72,Blondin-et-al-98}
\begin{equation}
 t\rs{tr}=2.83\E{4}E_{51}^{4/17}n\rs{H}^{-9/17}\un{yrs}
\label{rad-ttr}
\end{equation}
The radiative stage changes the post-adiabatic one around the shell-formation time \citep{Cox-Anderson-82,Cox-86}
\begin{equation}
 t\rs{sf}=5.18\E{4}E_{51}^{4/17}n\rs{H}^{-9/17}\un{yrs}.
\label{rad-tsf}
\end{equation}
These reference times are $t\rs{tr}=3.10\E{4}\un{yrs}$ and $t\rs{sf}=5.68\E{4}\un{yrs}$ for adopted $n\rs{H}$ and $E\rs{51}$. The duration of the post-adiabatic stage is $t\rs{sf}-t\rs{tr}=25800\un{yrs}$. 

The accuracy of our simulations is demonstrated by Fig.~\ref{tk-fig5bl}b: the total energy as a sum of components $E\rs{k}+E\rs{th}+E\rs{rad}$ is equal to the explosion energy up to the final time of simulations. (All these energies are calculated as an integrals of the respective local values over the whole spherical SNR volume.) 
The same plot shows that the energy of radiative losses increases, the thermal energy decreases while the kinetic energy is almost constant with time. Thus, the radiative losses affect mostly the thermal component of the total energy, in a way that the sum $E\rs{th}+E\rs{rad}$ is approximately constant in time. 

The semi-analytical approximate method for HD of the post-adiabatic SNR is developed by \citet{Hn-Pet-Tel-2007}. It uses Lagrangian coordinates and approximately reproduces the main features of the numerical simulations. 

\section{Role of magnetic field in evolution of the post-adiabatic SNRs}
%
\subsection{Effect of magnetic field on the dynamics of the shock}
%

In order to reveal the role of MF in the post-adiabatic evolution of SNRs, simulations were done for the two orientations of the ambient MF, parallel and perpendicular to the shock normal, and for few values of the ISMF strength: $B\rs{o}=1\un{\mu G}$, $3\un{\mu G}$, $10\un{\mu G}$ and $30\un{\mu G}$. 
In the simulations, the shock position was determined from the condition of the maximum of $|\nabla p|/p$ where $p$ is the thermal pressure.

\begin{figure}
\centering 
\includegraphics[width=8truecm]{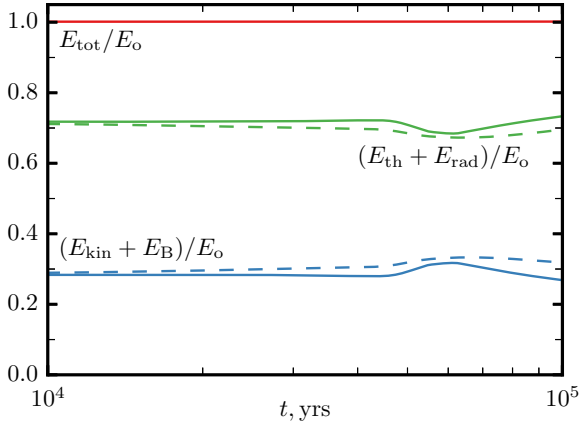}
\caption{Time dependence of the sums $E\rs{th}+E\rs{rad}$ and $E\rs{kin}+E\rs{B}$ in the simulations for $B\rs{o\perp} = 10\un{\mu G}$ (solid lines) and $B\rs{o\perp} = 30\un{\mu G}$ (dashed lines).}
\label{fig_energysums}
\end{figure}

\begin{figure*}
\centering 
\includegraphics[width=0.98\textwidth]{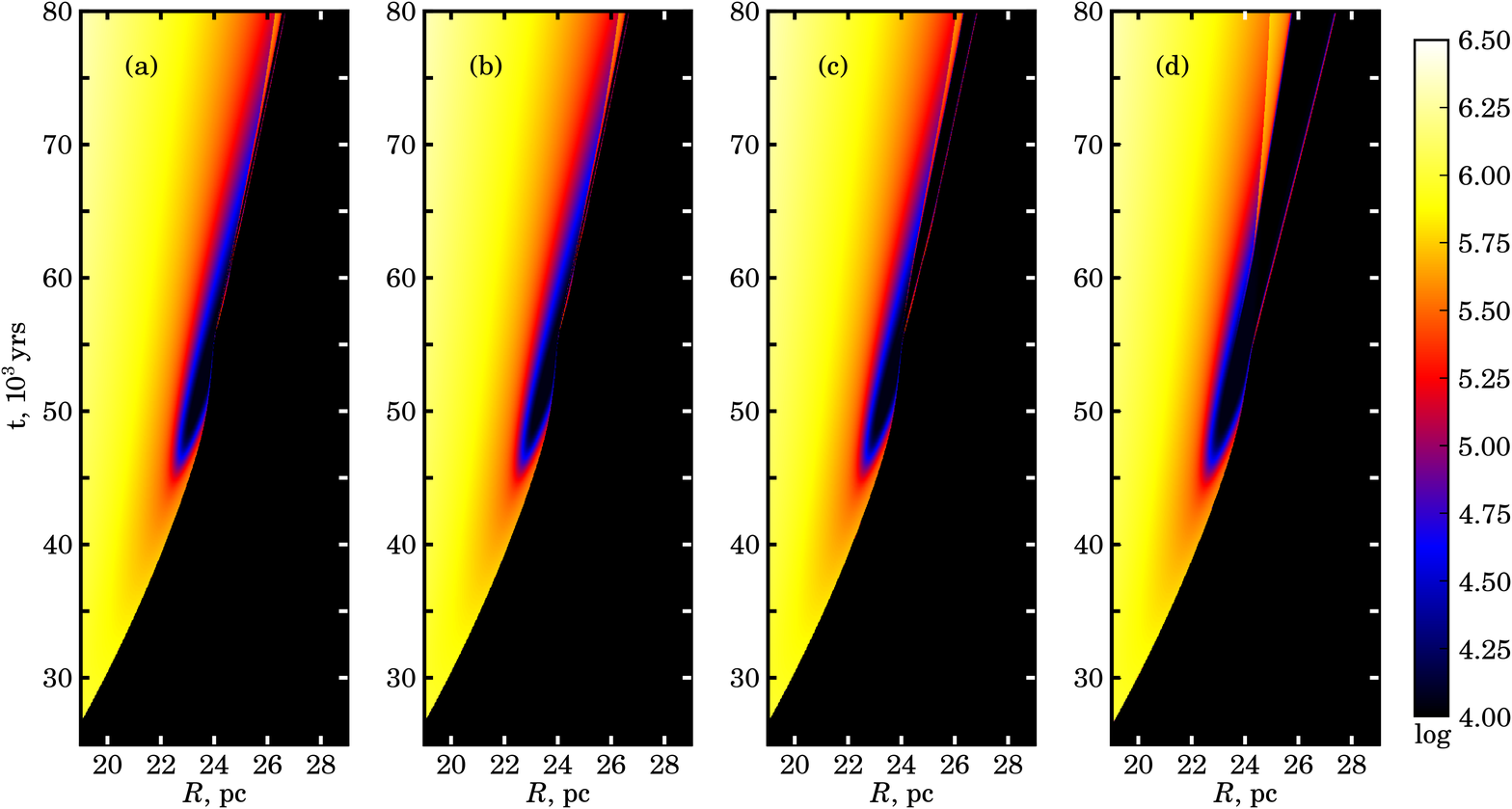}
\caption{Time dependence of the distribution of temperature 
behind the shock.
({\bf a}) $B\rs{o} = 0\un{\mu G}$, ({\bf b}) $B\rs{o\parallel} = 3\un{\mu G}$, 
({\bf c}) $B\rs{o\perp} = 3\un{\mu G}$, ({\bf d}) $B\rs{o\perp} = 10\un{\mu G}$.
Color scale is the same for all plots (in Kelvin, logarithmic scale).}
\label{fig_space_time_T}
\end{figure*}

Our simulations reveal that MF does not affect dynamics of the shock if it moves along the MF lines. The reason is that the energy density of the radial MF is much smaller than the thermal energy density. The situation is different for the perpendicular shock. The strength of the tangential MF increases considerably after passage of the shock front. 

Fig.~\ref{fig_blast_dynamics}a shows the evolution of the shock radius $R$ and velocity $V$ for the parallel and perpendicular orientations of ISMF. If the ambient MF is quite high then it is even possible that dynamics of the SNR radius is close to the extrapolation of the Sedov formula (Fig.~\ref{fig_blast_dynamics}a, orange dashed line),  despite the fact that radiation losses are significant (Fig.~\ref{fig_blast_dynamics}d, orange line). Effect of the strong MF is visible also on the plots for the shock velocity $V$ (Fig.~\ref{fig_blast_dynamics}a) and the deceleration parameter $m$ (Fig.~\ref{fig_blast_dynamics}c): the shock decelerates slower for higher MF strengths; MF works against of oscillations of $m$. In addition, the value of $m$ on the radiative stage ($t>t\rs{sf}$) is higher than the value expected from the pure hydrodynamical model $m\approx 0.3$ \citep{Band-Pet-2004} and could be close to the adiabatic value $0.4$ for large $B$ (Fig.~\ref{fig_blast_dynamics}b, orange dashed line). 

Fig.~\ref{fig_blast_dynamics}с and d demonstrate evolution of the components of the total energy for the perpendicular shock with $B\rs{o\perp} = 10\un{\mu G}$ and  $B\rs{o\perp} = 30\un{\mu G}$ respectively. The magnetic energy is calculated on the plots as 
\begin{equation}
E\rs{B}=\int_{V}\frac{B^2}{8\pi}\,dV-\int_{V}\frac{B\rs{o}^2}{8\pi}\,dV,
\end{equation}
i.e. it is the energy transformed to MF from the explosion energy.\footnote{The same approach was used for the thermal energy $E\rs{th}$: it is calculated with $(p-p\rs{o})$.} 

Magnetic energy is small for strengths $B\rs{o\perp}\leq 1 \un{\mu G}$ and dynamics of the energy components is similar to behavior in simulations (Fig.~\ref{tk-fig5bl}) without MF. 

The evolution of $E\rs{kin}$ and $E\rs{th}$ in the presence of the only parallel MF is the same as in purely HD simulations, for different strengths $B\rs{o}$. 

Radiative losses affect mostly the thermal energy, even if MF is strong (Fig.~\ref{fig_blast_dynamics}c-d). Some decrease of kinetic energy with time is visible; it transforms into the energy of the perpendicular MF. 
Thus, we observe almost independent channels of energy transformations: the thermal goes into the radiative and the kinetic into the MF energy. This statement is illustrated clearly on Fig.~\ref{fig_energysums}: the sums $E\rs{th}+E\rs{rad}$ and $E\rs{kin}+E\rs{B}$ are approximately constant with time. Since radiative losses come mostly from the thermal energy, the expressions (\ref{rad-ttr}) for the transition and (\ref{rad-tsf}) for the shell formation times are valid also in MHD models. 

MF energy increases with time and may reach prominent fraction of the total energy, e.g. for $B\rs{o\perp}=30\un{\mu G}$, it is about $0.2E\rs{tot}$ at $t=100\, 000\un{yrs}$ (Fig.~\ref{fig_blast_dynamics}d). Comparison of Fig.~\ref{tk-fig5bl}c, \ref{fig_blast_dynamics}c and d reveals that radiative losses are somehow smaller for stronger MF. In fact, at $t=100\, 000\un{yrs}$, about 60 \% of the explosion energy are lost through the thermal radiation if the shock moves in ISM without MF or along the MF lines (Fig.~\ref{tk-fig5bl}c), whereas the fraction of radiated energy is 
50 \% for $B_{0\perp} = 30\un{\mu G}$ (Fig.~\ref{fig_blast_dynamics}d).

\begin{figure*}
\centering 
\includegraphics[width=0.98\textwidth]{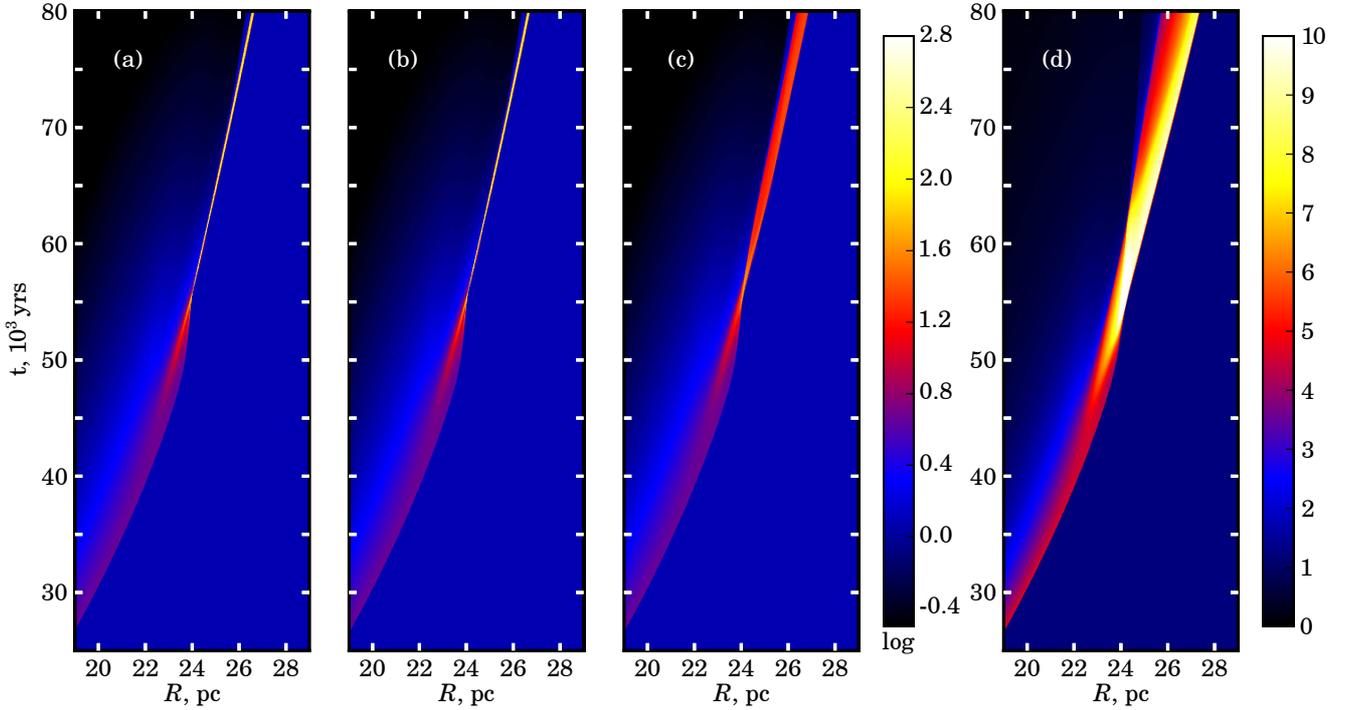}
\caption{Time dependence of the distribution of the number density 
behind the shock.
({\bf a}) $B\rs{o} = 0\un{\mu G}$, ({\bf b}) $B\rs{o\parallel} = 3\un{\mu G}$, 
({\bf c}) $B\rs{o\perp} = 3\un{\mu G}$, ({\bf d}) $B\rs{o\perp} = 10\un{\mu G}$.
Color bar on the right of the plot ({\bf c}) is for the three plots on the left 
(in $\un{cm^{-3}}$, logarithmic scale). 
Color bar on the right corresponds to the plot ({\bf d}) 
(in $\un{cm^{-3}}$, linear scale).}
\label{fig_space_time_rho}
\end{figure*}

%
\subsection{Effect of magnetic field on hydrodynamic parameters of the flow}
%

Fig.~\ref{fig_space_time_T} and \ref{fig_space_time_rho} shows the time evolution of the spatial distributions of the flow parameters in the region $19 - 29\un{pc}$ and in the time interval $25\,000 - 80\,000\un{yrs}$. 

The appearance of the first flow elements which radiated their energy away may clearly be tracked on the spatial-temporal distributions of temperature (Fig.~\ref{fig_space_time_T}). Radiative losses lead to the quick fall of the temperature behind the shock starting from $t\approx 40\,000\un{yrs}$. 
The first elements to cool appear a bit downstream of the shock, in agreement with \citet{Blondin-et-al-98}. 
If MF is low, radiative losses are effective in the thin region behind the shock. This region corresponds to the dense shell (\ref{fig_space_time_rho}a-b) which cools more and more effectively with increase of its density.

\begin{figure}
\centering 
\includegraphics[width=8.7truecm]{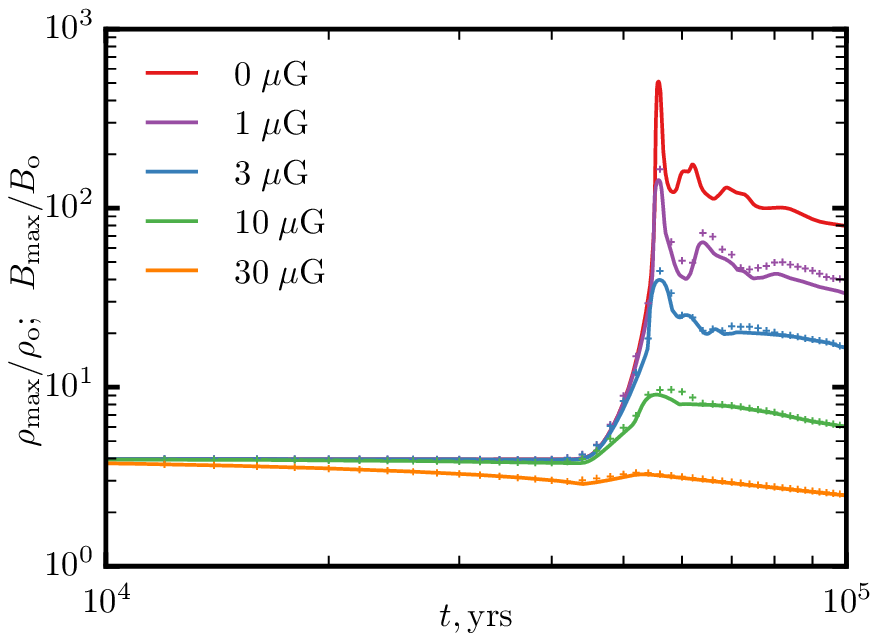}
\caption{Time dependence of the maximum density in the shell $\rho\rs{max}/\rho\rs{o}$ (solid lines) as well as the maximum value of MF strength $B\rs{max}/B\rs{o}$ (dotted lines), for few values of the strength of the ambient perpendicular MF.}
\label{fig_rho_max_log}
\end{figure}

The shell is more extended for $B\rs{o\perp} = 3\un{\mu G}$ (Fig.~\ref{fig_space_time_rho}c). 
The thin dense shell is not formed for $B\rs{o\perp} = 10\un{\mu G}$ (Fig.~\ref{fig_space_time_rho}d).
This is because the magnetic pressure is high enough (comparing to simulations with small MF strength) to prevent the huge compression of gas on the post-adiabatic and radiative phases. 
Fig.~\ref{fig_space_time_rho}a-b shows that the maximum density may be few hundred times the density in ISM, if MF is zero or parallel to the shock normal. 
In contrast, Fig.~\ref{fig_space_time_rho}c demonstrates that the compression factor of the perpendicular shock with $B\rs{o\perp} = 3\un{\mu G}$ is smaller than 50. The density compression factor for the post-adiabatic SNRs varies between 4 and 10 for $B_{0\perp} = 10\un{\mu G}$ (Fig.~\ref{fig_space_time_rho}d). 

This effect is visible more clearly on Fig.~\ref{fig_rho_max_log} where the evolution of $\rho\rs{max}/\rho\rs{o}$ and $B\rs{max}/B\rs{o}$ are shown for different values of perpendicular MF ($\rho\rs{max}$ and $B\rs{max}$ are the maximum values within the shell, its location within the shell varies with time). It is interesting to note on this figure that the shock compression is smaller than 4 for $B\rs{o}=30\un{\mu G}$; so large MF contributes to the ambient pressure and the shock is not `strong' any more. 

The smaller the density in the shell the smaller the radiative losses. On the other hand, the region affected by these losses are larger because the shell is thicker, if MF is accounted. In fact, it is prominent on Fig.~\ref{fig_space_time_T}b and Fig.~\ref{fig_space_time_T}c that 
the area behind the shock front, which is affected by the radiative losses, is wider in the case of perpendicular MF with $B\rs{o\perp}=3\un{\mu G}$, and increases with the value of the MF strength (Fig.~\ref{fig_space_time_T}d). 

Large value of the perpendicular MF creates an additional pressure immediately downstream which prevents the flow from large compression (which is present in the pure HD simulations). It also pushes the shock to the larger distance. 
This results in $R$, $V$ and $m$ being closer to the Sedov extrapolations, even in the radiative phase $t>t\rs{sf}$ 
(Fig.~\ref{fig_blast_dynamics}a-b, green and orange lines). In this respect, MF behaves like a `compensator' of pressure which is lost due to radiative losses.

Fig.~\ref{fig_space_time_rho} confirms that the thickness of the shell for the large strength of the perpendicular MF is not small anymore and increases with time. Therefore, the theoretical model of the `snowplow' may not be used. 

\begin{figure*}
\centering 
\includegraphics[width=0.8\textwidth]{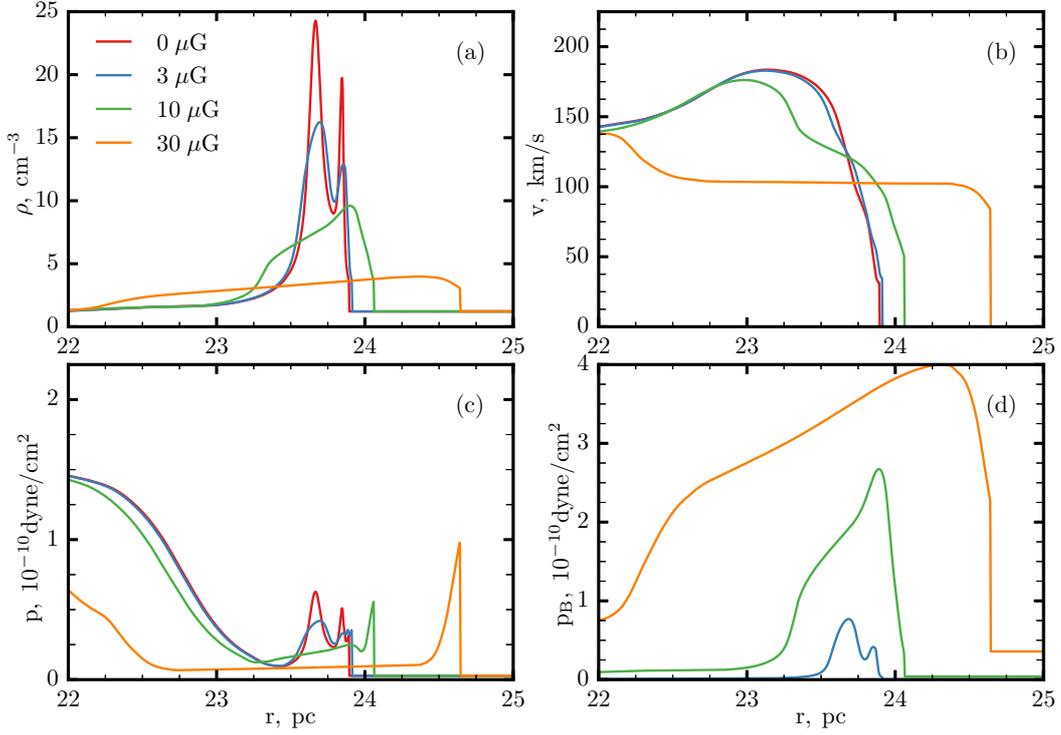}
\caption{Spatial distribution of parameters downstream of the perpendicular shock for $t = 53\, 000$ yrs: 
({\bf a}) density, ({\bf b}) velocity, ({\bf c}) thermal pressure, ({\bf d}) MF pressure. 
Distributions behind the parallel shock coincide with the results for $B\rs{o} = 0\un{\mu G}$.}
\label{fig_plot_profile_530}
\end{figure*}
\begin{figure*}
\centering 
\includegraphics[width=0.8\textwidth]{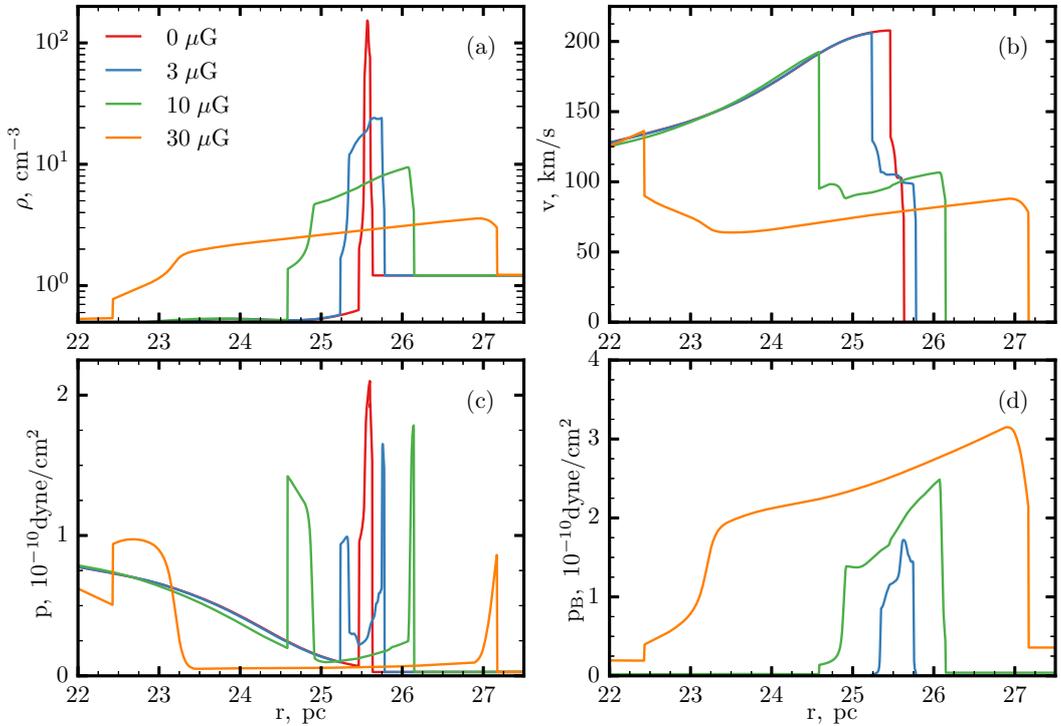}
\caption{The same as on Fig.~\ref{fig_plot_profile_530} for $t = 70\, 000$ yrs.}
\label{fig_plot_profile_700}
\end{figure*}
\begin{figure*}
\centering 
\includegraphics[width=0.8\textwidth]{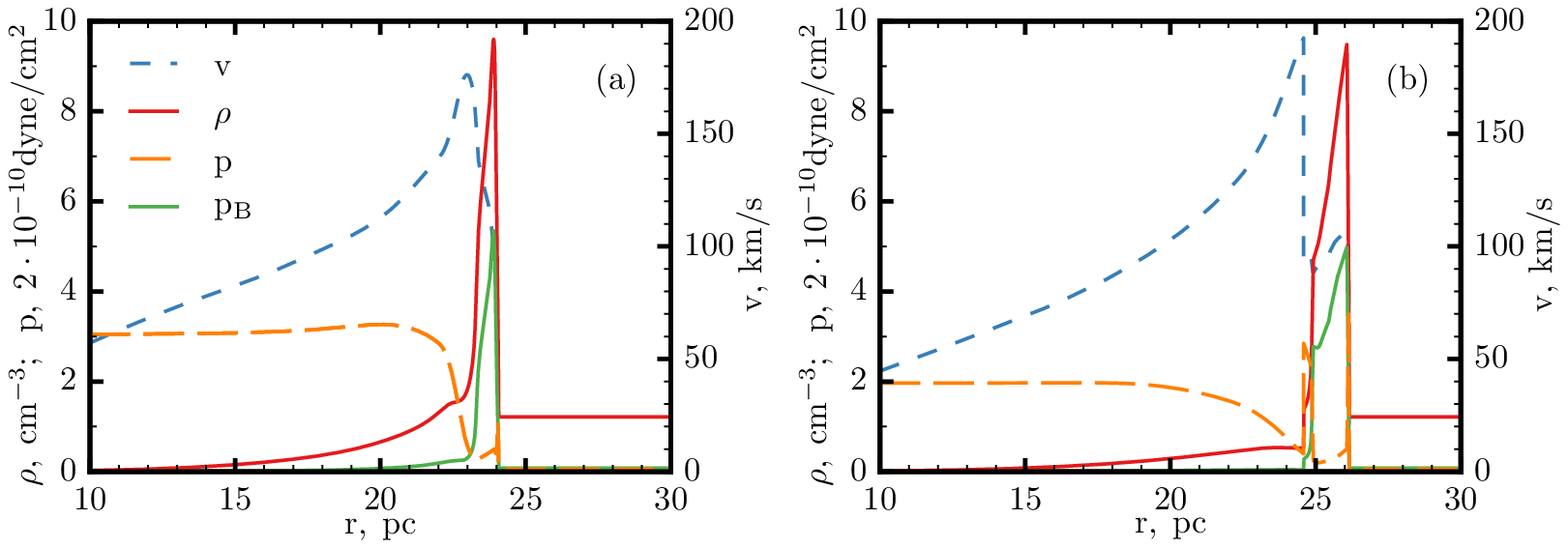}
\caption{Radial profiles of the velocity $v$, density $\rho$, 
the thermal pressure $p$ and the magnetic pressure $p\rs{B}$ for 
$B_{o\perp} = 10\un{\mu G}$ and the SNR age of ({\bf a}) $53\, 000\un{yrs}$, ({\bf b}) $70\, 000\un{yrs}$.
}
\label{fig_plot_full}
\end{figure*}

It is known from the hydrodynamic simulations \cite[e.g.][]{Blondin-et-al-98} that, at the post-adiabatic stage, the two shells form (Fig.~\ref{fig_plot_profile_530}). They merge into one around time ($t \approx 55\, 000\un{yrs}$) which corresponds to the first minimum of the deceleration parameter $m$ (Fig.~\ref{fig_blast_dynamics}c); 
this is visible on Fig.~\ref{fig_space_time_rho} for $B_{o} = 0\un{\mu G}$ and $B\rs{o\parallel} = 3\un{\mu G}$. 
Two shells are formed because the effective cooling happens and develops a bit downstream. The pressure decrease in the outer shell results in the shock deceleration; this is reflected by the decrease of the parameter $m$ up to the first minimum (Fig.~\ref{fig_blast_dynamics}b). The internal pressure pushes the inner shell until both shells merge (time of the minimum of $m$) and then it pushes the merged shell (the first increase of $m$, Fig.~\ref{fig_blast_dynamics}b). The shell experiences the radiative losses and slows down (the second decrease on $m$ around $t=6\cdot 10^4\un{yrs}$) reducing the ram pressure of the incoming gas. Following oscillations of $m$ are similar, they are due to the interplay between the internal and the ram pressures regulated by the radiative losses. The overall evolution of the structures due to the radiative cooling, in particular, the formation of secondary shocks, are similar to described in \citet{Innes-1992}.

Perpendicular MF introduces the additional, magnetic, pressure in this picture. 
When $m$ approaches its first minimum, the magnetic pressure becomes comparable with the thermal one. 
Later, MF appears to be an intermediate chain in interactions between the internal and the ram pressures: the two peaks of the pressure are clearly separated by the MF pressure (Fig.~\ref{fig_plot_profile_700}c; Fig.~\ref{fig_plot_full}). This effect is visible after $t\approx 55\,000\un{yrs}$ (which corresponds to the first minumum of $m$) for $B\rs{o\perp}=3\un{\mu G}$ (cf. Fig.~\ref{fig_plot_profile_530} and \ref{fig_plot_profile_700}). Larger MF prevents the two shells from interactions during the whole post-adiabatic phase, even around $t\approx 55\,000\un{yrs}$ (Fig.~\ref{fig_space_time_rho}d). 

\begin{figure*}
\centering 
\includegraphics[width=0.8\textwidth]{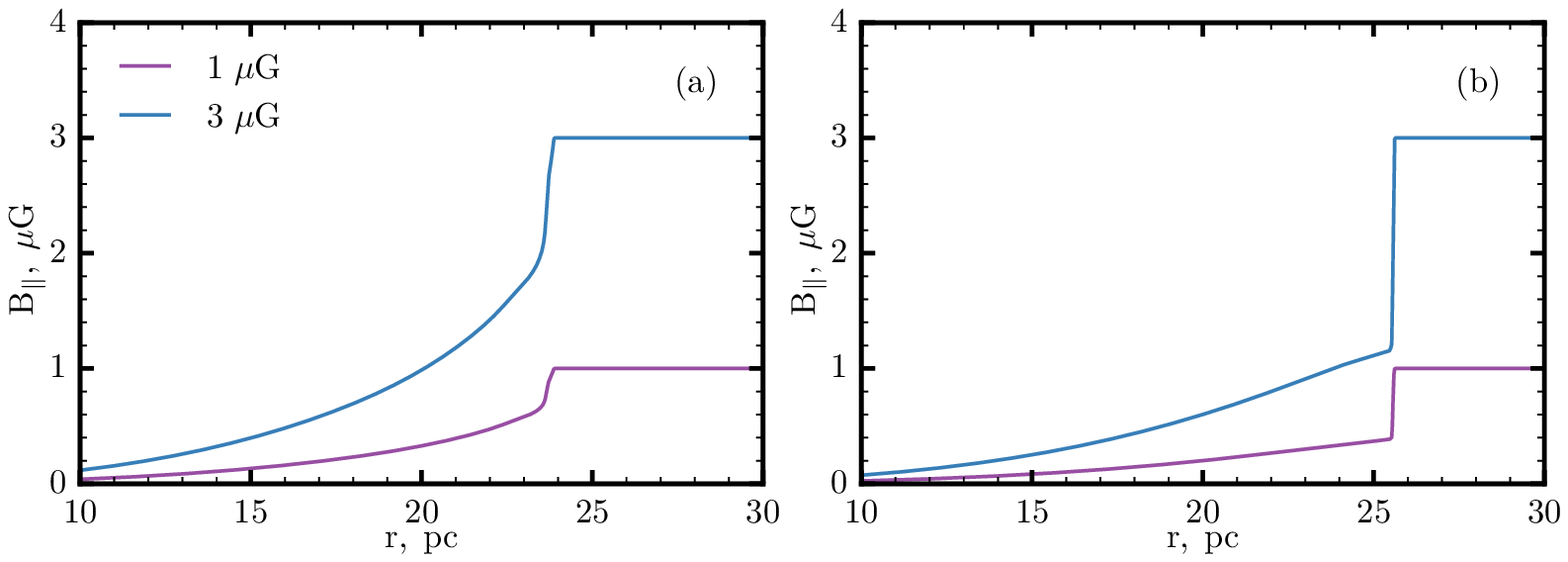}
\caption{Radial profiles of the magnetic field downstream of the parallel shock 
for SNR age $t = 53\, 000$ yrs ({\bf a}), $t = 70\, 000$ yrs ({\bf b}) and 
two values of $B\rs{o\|}$.}
\label{fig_mf_parallel}
\end{figure*}
\begin{figure}
\centering 
\includegraphics[width=8truecm]{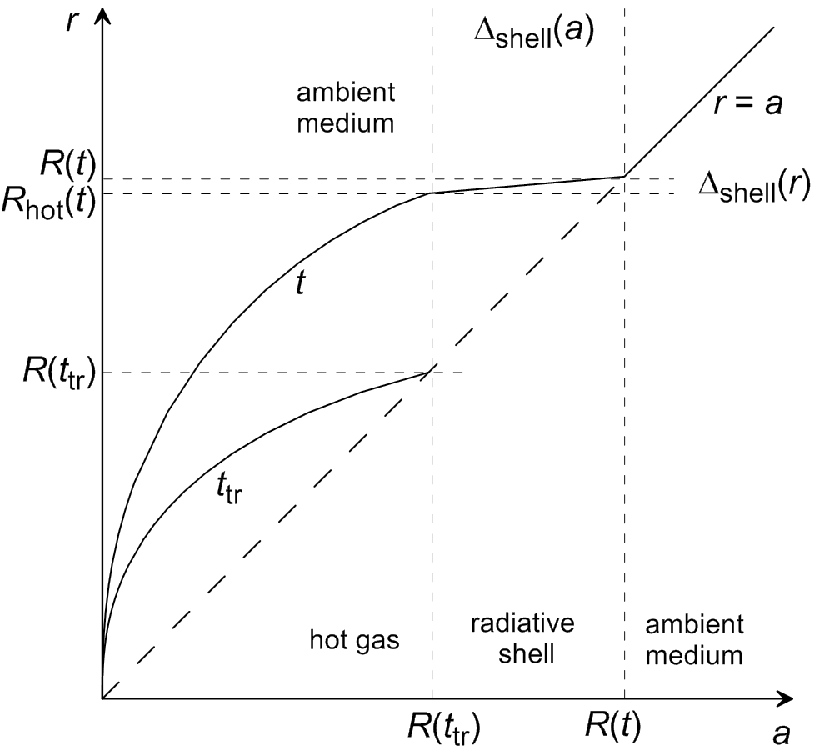}
\caption{Schematic representation of the relation between the Lagrangian $a$ and Eulerian $r$ coordinates.}
\label{fig_sxema}
\end{figure}

Figs.~\ref{fig_plot_profile_530}, \ref{fig_plot_profile_700} and \ref{fig_plot_full} correspond to the time before and after the first minimum of the deceleration parameter $m$.
The figures demonstrate that the thickness of the whole radiative shell increases with increase of the strength of ambient MF. MF reduces the density in the shell; the density is smaller for larger MF strength (Fig.~\ref{fig_rho_max_log}). 
The shock goes forward to the larger extent if tangential MF is present; this effect is more prominent for higher $B\rs{o\perp}$ (Fig.~\ref{fig_blast_dynamics}, \ref{fig_plot_profile_530}a, \ref{fig_plot_profile_700}a).


Simulations with parallel orientation of MF show no influence of MF on the distribution of hydrodynamical parameters of the flow. The distributions are the same as in the purely hydrodynamical simulations. 

%
\subsection{Structure of magnetic field}
%

Fig.~\ref{fig_mf_parallel} shows distributions of MF strength for $B_{o\parallel} = 1\un{\mu G}$ and $B_{o\parallel} = 3\un{\mu G}$ for $t= 53\, 000\un{yrs}$ and $70\,000\un{yrs}$.
The parallel MF is not compressed behind the shock, as expected (Fig.~\ref{fig_mf_parallel}). 
MF distribution is self-similar in respect of the MF strength, i.e. the MF profiles $B\rs{\|}(r)/B\rs{o\|}$ are the same for different values of $B\rs{o\|}$ but the shape of the profile varies in time. After 
$t\approx 50\,000\un{yrs}$, there is a significant drop in the magnetic field right after the shock; the drop  increases with time. In particular, at $t=10^5\un{yrs}$, the MF strength becomes $0.25 B\rs{o\|}$. 

Such a behavior is related to the formation of the thin dense shell. Fig.~\ref{fig_sxema} is useful for understanding. 
It shows the relation between the Lagrangian $a$ and Eulerian $r$ coordinates. The coordinate $r$ is common spatial position. The coordinate $a$ is like a number attached to the fluid element; it is defined as $a=R(t\rs{i})$, i.e. it is the shock radius at time $t\rs{i}$ when the given fluid element was shocked. Therefore, $a=r$ in the ambient medium. During the adiabatic stage the profile $r(a)$ is self-similar, i.e. it has the same shape for any time. The  shell starts to form approximately at $t\rs{tr}$. 
The radiative shell consists mostly of the swept up ISM\footnote{The minor contribution to the shell comes also from the cooled material of the SNR interior, within about $5\%$ of the radius $R(t\rs{tr})$ \citep{Blondin-et-al-98}; this contribution is neglected on the Fig.~\ref{fig_sxema}.} \citep[e.g.][]{Blondin-et-al-98}. Therefore, the shell is thick in coordinates $a$; the thickness is $\Delta\rs{a}=R(t)-R(t\rs{tr})$ and increases with time. The hot gas behind the shell (with $a<R(t\rs{tr})$) evolves almost adiabatically, thus the relation $r(a)$ for this region  scales with time approximately as $r(a,t)\approx r(a,t\rs{tr})R\rs{hot}(t)/R(t\rs{tr})$ (see Fig.~\ref{fig_sxema}). The pressure in the shell falls due to the radiative losses; therefore, the gas with Lagrangian coordinates $R(t\rs{tr})<a<R(t)$ is compressed in a dense shell which is quite thin in common coordinates $r$: $\Delta\rs{r}=R(t)-R\rs{hot}(t)$. 

The MF components behave in accordance to the fundamental Eqs.~(\ref{rad:MFlagr}) which are valid either for adiabatic or for non-adiabatic regime. Let's consider a fluid element which is located downstream but quite close to the shock, say $r=R-\Delta\rs{r}$. The coordinate $a$ which corresponds to this $r$ is considerably smaller: $a=R-\Delta\rs{a}$ (see Fig.~\ref{fig_sxema}). In other words, the ratio $a/r$ is smaller than unity and decreases with time. This is why $B\rs{\|}=(a/r)^2$ experiences rapid fall behind the shock, as on Fig.~\ref{fig_mf_parallel}. Note, that this effect also explains why $B\rs{max}/B\rs{o}$ is typically larger than $\rho\rs{max}/\rho\rs{o}$ on Fig.~\ref{fig_rho_max_log}: the perpendicular component of MF $B\rs{\perp}/B\rs{o}=(r/a)(\rho/\rho\rs{o})$. 

\begin{figure*}
\centering 
\includegraphics[width=0.98\textwidth]{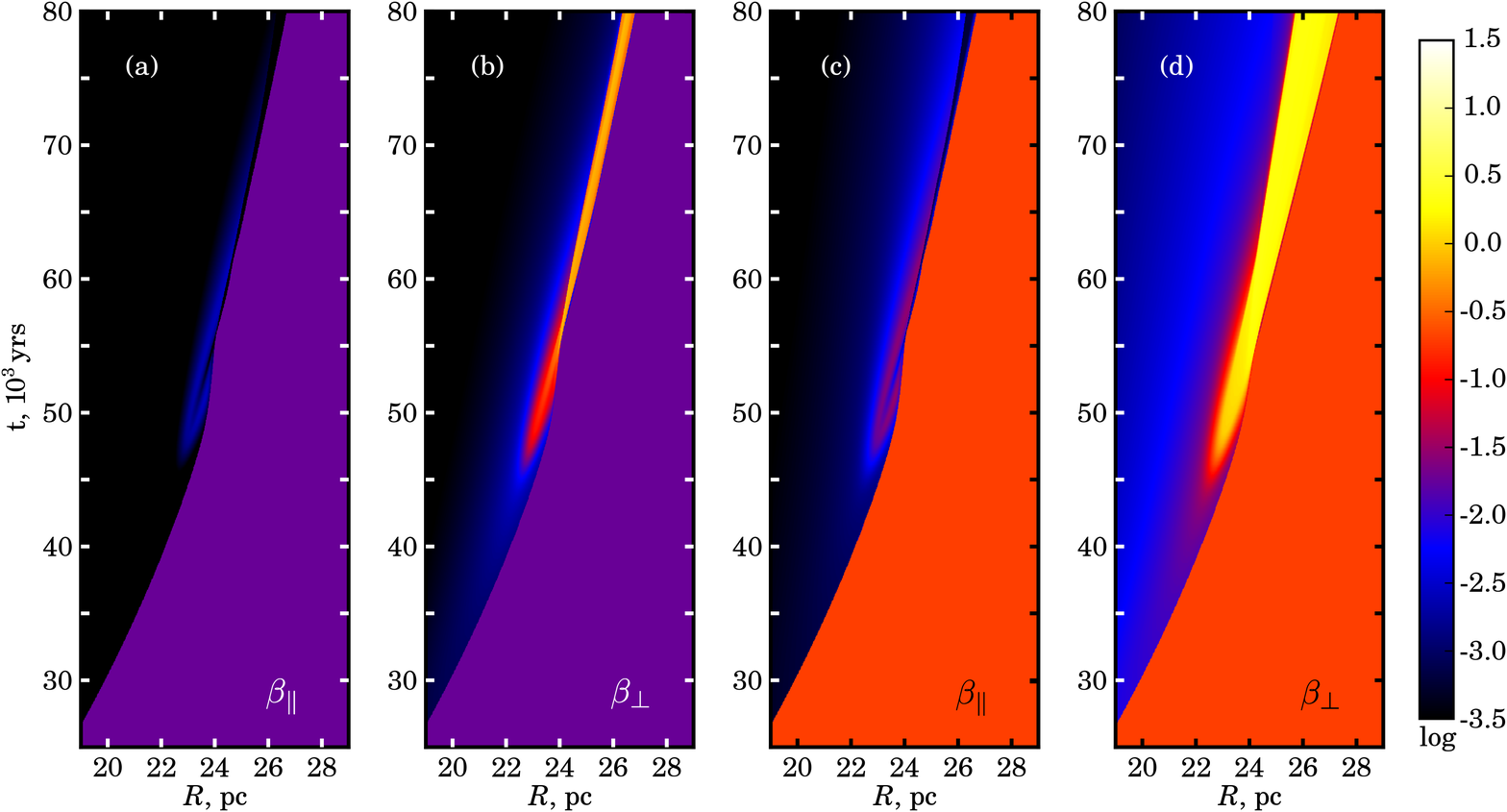}
\caption{Time dependence of the distribution of the plasma $\beta$; 
it is a ratio between the energy densities of MF and gas. 
({\bf a}) $B\rs{o\|} = 3\un{\mu G}$, ({\bf b}) $B\rs{o\perp} = 3\un{\mu G}$, 
({\bf c}) $B\rs{o\|} = 10\un{\mu G}$, ({\bf d}) $B\rs{o\perp} = 10\un{\mu G}$.
Color scale is logarithmic and the same for all plots.}
\label{fig_plasma_beta}
\end{figure*}
\begin{figure*}
\centering 
\includegraphics{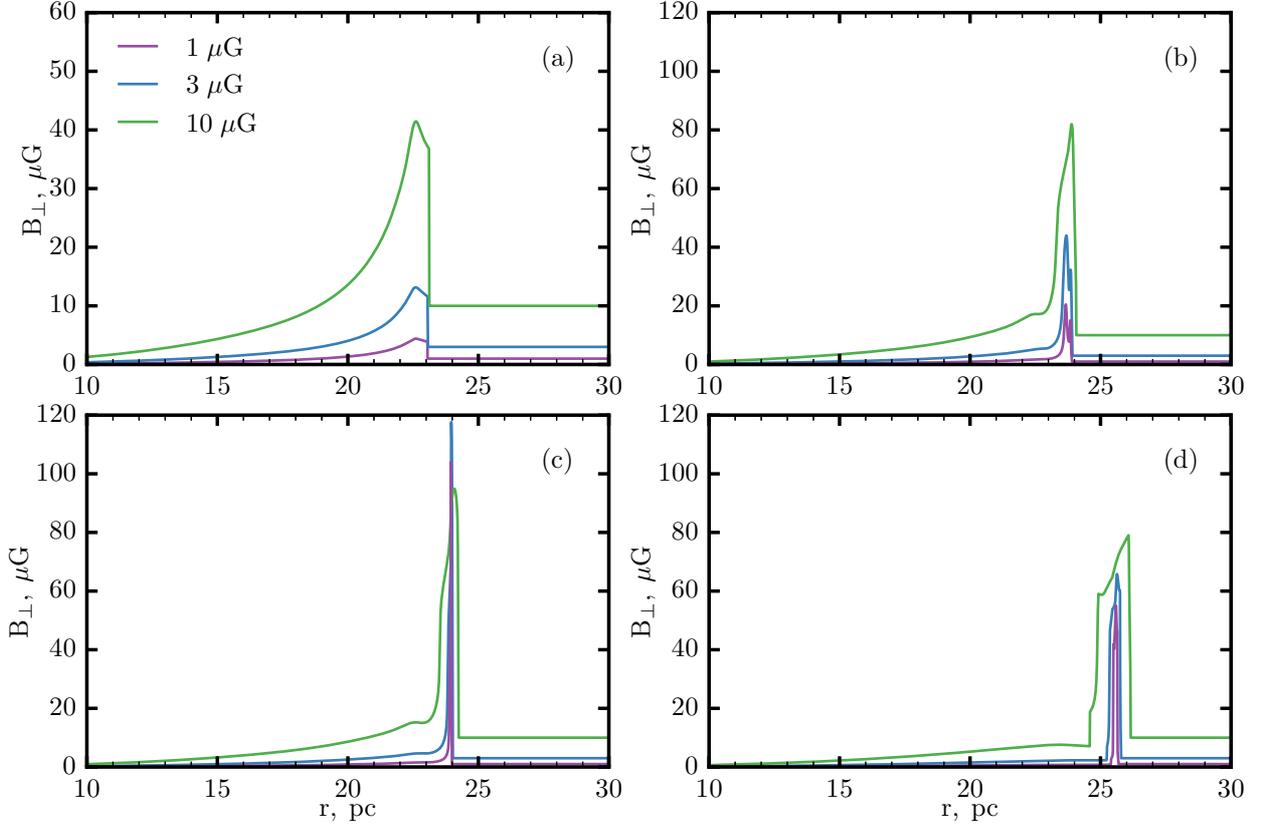}
\caption{Radial profiles of the magnetic field downstream of the perpendicular shock for
$t = 45\, 000$ yrs ({\bf a}), $t = 53\, 000$ yrs ({\bf b}), $t = 55\, 000$ yrs ({\bf c}), $t = 70\, 000$ yrs ({\bf d}) 
and few values of $B\rs{o\perp}$.}
\label{fig_mf_perpendicular}
\end{figure*}

The component of MF parallel to the shock normal does not affect the flow structure during the whole evolution of SNR. The parallel field depends on the hydrodynamic structure rather than vice versa. 
Fig.~\ref{fig_plasma_beta}a,c explains this effect: the plasma $\beta$ is less than unity everywhere inside SNR. 

Spatial distributions of the perpendicular MF (Fig.~\ref{fig_mf_perpendicular}) correlate with distributions of  density (Fig.~\ref{fig_plot_full}) because MF is frozen into the plasma. At the beginning of the post-adiabatic stage,  when the radiative losses are not essential yet, the distributions for MF of the intermediate strength ($B\rs{o\perp}=1\un{\mu G},\ 3\un{\mu G}$) are close to the self-similarity in respect of the value of $B\rs{o}$ (like in the case of the parallel field). However, this property is quickly violated due to different behavior of the radiative loses in the flows modified by MF of the different strengths. The tangential MF is important factor in dynamics of the flows with shocks, in accordance with the values of the plasma $\beta$ (Fig.~\ref{fig_plasma_beta}b,d). 

The strength of the perpendicular MF may reach quite high values on the post-adiabatic and radiative stages, even for small values $B\rs{o\perp}\sim 1\un{\mu G}$ because if $\beta$ is small then MF does not prevent plasma from high compression and the perpendicular MF is compressed to similar values as density. MF with higher $B\rs{o\perp}$ limits the compression and the maximum value of the strength is smaller (Fig.~\ref{fig_rho_max_log}).

\section{Conclusions}

The numerical code PLUTO was adopted and used for magneto-hydrodynamic simulations of flows with shock waves in supernova remnants with the radiative losses of plasma. Calculations are performed for the case of the one-dimensional flows under conditions of spherical symmetry, in the interstellar medium with the uniform distributions of density and  magnetic field. The role of MF in formation of the flow structure is revealed. It is significant only in the post-adiabatic and radiative phases.

The pressure of the tangential MF prevents large compression of plasma and limits the level of maximum density in the radiative shell. Radiative shell is more extended behind the perpendicular shock. These effects are more significant for larger values of MF strength.

Parallel MF does not affect the flow dynamic because its energy density is much smaller than the thermal energy density. 

Radiative losses affect mostly the thermal energy. The magnetic energy increases mostly by reducing the kinetic energy. 

At oblique shocks of the post-adiabatic SNRs, the perpendicular component of MF increases and the parallel component decreases. This mechanism may be responsible for the predominantly tangential orientation of MF in old SNRs, as it is known from observations of the radio polarization \citep{Milne-1987,Dubner-Giacani-2015}. 

At late times, when the shock decelerates a lot and the Mach number becomes small, efficiency of particle acceleration by the Fermi mechanism considerably decreases. On the other hand, at these times the tangential MF dominates over most of the SNR surface, even in the regions where the parallel MF was large at previous stages (quasiparallel shocks). This probably keeps more or less effective further acceleration of particles due to the drift mechanism. Especially in the view of the rather small shock velocity: the particles tied to MF lines might spend more times for multiple crossings of the (quasi-)perpendicular shock until the MF line will be swept up downstream. 

The old post-adiabatic SNRs or regions of SNRs interacting with the high-density medium may be promising sources of hadronic \g-rays. Radiative losses in these objects essentially affect dynamics of the shock and structure of the flow. They cause considerable increase of density and MF strength in the radiative shell. Therefore, prominent emission of hadronic \g-rays should be generated there whereas the competing leptonic \g-rays has to be substantially reduced under such conditions. 

\section*{Acknowledgements}

All simulations were performed on the computational cluster in Institute for Applied Problems in Mechanics and Mathematics. The research was partially supported by the Program of Ukrainian National Academy of Sciences "Grid infrastructure and grid technologies for science and applications" (grant 0115U002936).





\bsp
\label{lastpage}
\end{document}